\documentclass[aps,twocolumn,a4paper,showpacs, superscriptaddress]{revtex4}
\usepackage{graphicx}
\usepackage[all]{xy}
\usepackage{amsmath}
\usepackage{amssymb}
\usepackage{color}

\newcommand{\bb}{\bibitem}
\newcommand{\bes}{\begin{subequations}}
\newcommand{\ees}{\end{subequations}}
\def\ben{\begin{eqnarray}}
\def\een{\end{eqnarray}}
\def\be{\begin{equation}}
\def\ee{\end{equation}}

\begin{document}
\title{A Note On Quantum Compactons} 
\author{D. Bazeia} 
\affiliation{Departamento de F\'\i sica Universidade Federal da Para\'iba, 58051-900 Jo\~ao Pessoa PB, Brazil}
\author{D.V. Vassilevich}
\affiliation{Centro de Matem\'atica, Computa\c c\~ao e Cogni\c c\~ao, Universidade Federal do ABC, 09210-170 Santo Andr\'e, SP, Brazil}
\begin{abstract}
Compactons are solutions of the equations of motion that behave trivially outside a compact region. In general, the operators describing quantum fluctuations above compactons have singularities. However, we show that despite these singularities the quantum theory is well defined. As an example, we calculate the one-loop mass shift of a compacton in a model described by a single real scalar field. 
\end{abstract}
\maketitle

\section{Introduction}

Topological defects appear in nature in a diversity of contexts. In particular, they are of current interest in high energy physics, where they can be used to describe phase transitions in the early universe or map interfaces separating distinct regions in space \cite{r1,VS,MS}. An interesting recent example of the use of topological defect is the study of magnetic domain wall in a nanowire, intended for the development of magnetic memory at the nanometric scale \cite{AV}. Other investigations concerning topological defects in high energy physics can be found for instance in \cite{DM}.

In this work we study the presence of topological defects in relativistic models described by a real scalar field $\phi$ in $(1+1)$ spacetime dimensions. We focus attention on k-field models, with the kinematics modified to allow for higher order power in the derivative of the dynamical field. These models were introduced to contribute to understand the current accelerated expansion of the Universe \cite{K1}, and they were also studied with other motivations in Ref.~\cite{K2,CoA1,CoA2,CoB}, for instance. See also \cite{BA} for supersymmetric extensions of these models. 

Our main focus is on the compacton solutions in models engendering
generalized kinematics \cite{CoA1,CoA2,CoB}. We remind that compactons appear in the presence of nonlinearity and nonlinear dispersion, acquiring spatial profile with compact support \cite{RH}. They have been investigated in different contexts in \cite{comp1,comp2} and they are spacelike structures similar to kinks; see, e.g. \cite{comp3} for a recent study on compactons.

Compactons behave trivially outside a compact region. Consequently, the derivatives may develop discontinuities, and the fluctuation operators may have singularities reminding confining domain walls. We study whether despite their exotic properties these solutions
can be quantized. We find that the answer is positive and calculate the one-loop shift of the compacton energy. 

We organize the work as follows: in Sec.~\ref{sec-cla} we briefly review the compacton classical solutions in a model with a single real scalar field in $(1+1)$ spacetime dimensions. In Sec.~\ref{sec-flu} we consider the quantum fluctuations and the radiative corrections. We end the work with some comments and conclusions in Sec.~\ref{sec-com}.

\section{Classical solutions}\label{sec-cla}

We consider the models described by the Lagrange density
\be
{\cal L}=F(X) - V(\phi),
\ee
where $V(\phi)$ is a potential depending on a real scalar field $\phi$. We work in $(1+1)$ spacetime dimensions, with $x^\mu=(x^0=t,x^1=x)$, $x_\mu=(x_0=t,x_1=-x)$ and we consider $t,x$ and the field $\phi$ dimensionless, for simplicity. The function $F(X)$ is in principle an arbitrary function of $X$, which is defined as
\be
X=\frac12\partial_\mu\phi\partial^\mu\phi.
\ee

The most famous model from this class with
\be
{\cal L}_0=X-\frac12 (1-\phi^2)^2.
\ee 
is the $\phi^4$ model, engendering spontaneous symmetry breaking. 
It supports defect structures, of the kinklike type, given explicitly by
\be
\phi_0(x)=\tanh(x).
\ee
Less standard models can be described by, for instance, 
\be
{\cal L}_n=-X^2-\frac34(1-\phi^2)^{2n}.
\ee
where $n$ can be 1 or 2, leading to the models
\bes\ben
{\cal L}_1=-X^2-\frac34(1-\phi^2)^{2};\label{n1act}\\
{\cal L}_2=-X^2-\frac34(1-\phi^2)^{4}.\label{n2act}\een\ees
The equation of motion has the general form
\be
\partial_\mu\left({\cal L}_X\partial^\mu\phi\right)+\frac{dV}{d\phi}.
\ee
Here we get
\be
\partial_\mu\phi\partial^\mu\phi\,\partial_\nu\partial^\nu\phi+2\partial_\mu\phi\partial_\nu\phi\,\partial^\nu\partial^\mu\phi=\frac{dV}{d\phi}.
\ee
In this case, for static solution the equations of motion are, for $n=1$
\be
\phi^{\prime 2}\phi^{\prime\prime}=-\phi(1-\phi^2),
\ee
and for $n=2$
\be
\phi^{\prime 2}\phi^{\prime\prime}=-2\phi(1-\phi^2)^3.
\ee
Also, the static solutions have energy density and stress given by
\bes\ben
\rho_n(x)=\frac14 \phi^{\prime 4}+V_n(\phi);
\\
\tau_n(x)=\frac34 \phi^{\prime 4}-V_n(\phi).
\een\ees

The two models  also support defect structures. For the model described by ${\cal L}_1$ we get
\be
\phi_1(x)=\sin(x),\;\;\; -\frac{\pi}{2}\leq x \leq \frac{\pi}2,\label{comp}
\ee
and $\phi_1(x)=-1$ for $x\leq -\pi/2$; $\phi_1(x)=1$ for $x\geq \pi/2$. The energy density is
\be
\rho_1(x)= \cos^4(x),
\ee
for $x\in [-\pi/2,\pi/2]$; $\rho_1(x)$ vanishes outside this compact region. The energy becomes $E_1=3\pi/8$. The solution is of the compact type, since it is defined in the real line, but it only deviates from the minima in the compact set $x\in[-\pi/2, \pi/2]$. We note that $\phi_1(x)=\sin(x)$ is stressless, leading to the first-order equation
\be
\phi^{\prime}=\sqrt{1-\phi^2}=W_{1\phi}^{1/3},
\ee
where
\be
W_1(\phi)=\phi\sqrt{1-\phi^2}\left( \frac58-\frac14 \phi^2\right)+\frac38 {\rm arcsin(\phi)}
\ee
and $W_{1\phi}=dW_1/d\phi$.

For the second model, described by ${\cal L}_2$, we get
\be
\phi_2(x)=\tanh(x)\,.\label{phi2}
\ee
This is the same solution we found in the standard $\phi^4$ model. See \cite{comp3} for further details on compactons.

\section{Fluctuations and radiative corrections}\label{sec-flu}
Let us study quantum fluctuation above the compacton solution \eqref{comp} in the model described by ${\cal L}_1$; see Eq.~\eqref{n1act}. Outside the region where this
solution is localized the dynamics of fluctuations is governed by the same action (\ref{n1act}) with a
shifted potential,
\begin{eqnarray}
&&\mathcal{L}(\eta)=-X^2(\eta)-\frac 34 (-\eta^2+2\eta)^2,\;\; x<-\pi/2;\\
&&\mathcal{L}(\eta)=-X^2(\eta)-\frac 34 (\eta^2+2\eta)^2,\quad x>\pi/2\,.\label{Lout}
\end{eqnarray}
Kinetic terms in both actions above are 4th order in fluctuations. Consequently, there are no
propagating perturbations outside the compacton. This effect was already noted in \cite{CoA2}.

Inside the region, $-\pi/2<x<\pi/2$, the linearized equations of motion lead to the following
Schr\"odinger-like equation 
\be\label{se}
Lu_\omega\equiv\left( -\frac{d^2}{dz^2}+ U_1(z)\right)u_{\omega}(z)=\omega^2 u_{\omega}(z),
\ee
where we have changed $x\to z$ and $\eta\to u$, according to
\be
z= 3^{1/2}x,\qquad
u_\omega=3^{1/4}\phi^{\prime}\,\eta_\omega. \label{ueta}
\ee
$U_1(z)$ is the P\"oschl-Teller potential \cite{PT}
\be
U_1(z)=-12+6\, \sec^2(\sqrt{3}z).\label{U1}
\ee
Solutions of the equation \eqref{se} are very well known, see \cite{PT,Fluegge65}:
\begin{eqnarray}
&&u_{2k}(z)=\cos^2(\sqrt{3}z)\, _2F_1(2+k,-k,\tfrac 12,\sin^2(\sqrt{3}z)),\nonumber\\
&&u_{2k+1}(z)=\cos^2(\sqrt{3}z)\sin(\sqrt{3}z)\nonumber\\
&&\qquad\qquad\qquad \times \, _2F_1(3+k,-k,\tfrac 32,\sin^2(\sqrt{3}z)),\label{uu}
\end{eqnarray}
$k=0,1,2,\dots$ The hypergeometric functions in the equations above are actually polynomials of 
$\sin^2(\sqrt{3}z)$ of degree $k$. The corresponding eigenfrequencies read
\begin{equation}
\omega_n=\sqrt{3(n^2+4n)}\,.\label{omegan}
\end{equation}

By looking at the functions \eqref{uu} we can make an important observation. They all vanish at least
as $\cos^2(\sqrt{3}z)$ at the endpoints of the interval $[-\pi/2\sqrt{3},\pi/2\sqrt{3}]$. Therefore,
the original perturbations $\eta$, see \eqref{ueta}, also vanish at these points. This implies that
quantum fluctuations cannot escape the region where the compacton is localized.

As an example of quantum computations with the compacton let us calculate the energy of confined
fluctuations. Formally, it is given by a half sum of the eigenfrequencies (\ref{omegan})
\begin{equation}
\mathcal{E}=\frac 12 \sum_{n=1}^\infty \omega_n\,. \label{Eom}
\end{equation}
This sum is obviously divergent and has to be regularized and renormalized. We use the zeta-function
regularization.  Earlier, the heat kernel methods and the zeta regularization were applied to calculations of the quantum energy of solitons in \cite{Bo,BGvNV}.
Namely, we introduce a complex parameter $s$ and replace \eqref{Eom}) by 
\begin{equation}
\mathcal{E}_s=\frac 12 \sum_{n=1}^\infty \bigl( \omega_n^2 \bigr)^{\frac 12 -s}\,. \label{Ezeta}
\end{equation}
For $\Re (s)>1$ the sum above converges, but gets divergent contributions (poles) when continued to the
physical value $s=0$. To analyze these divergences, it is convenient to make the Mellin transformation
\begin{equation}
\mathcal{E}_s=\frac 1{\Gamma \bigl( s-\tfrac 12 \bigr)} \int_0^\infty \frac {dt}{t} t^{s-\frac 12} K(t)\,,
\label{Ehk}
\end{equation}
where
\begin{equation}
K(t)=\sum_{n=1}^\infty e^{-t \omega_n^2}={\rm Tr}\left( e^{-tL}\right), \label{hk}
\end{equation}
is the heat kernel of the second-order differential operator $L$ defined in 
\eqref{se}. It is well known, that for $\tau\to +0$,
\begin{equation}
\sum_{k=-\infty}^\infty e^{-\tau k^2} \simeq \sqrt{ \frac {\pi}\tau} ,\label{Poisson}
\end{equation}
up to exponentially small terms. By using this formula, one can easily derive the small $t$
asymptotic expansion of the heat kernel (\ref{hk}):
\begin{eqnarray}
K(t)\simeq&& \frac 12 \sqrt{\frac{\pi}{3t}} -\frac 52 + 2\sqrt{3\pi t}+O(t),\nonumber\\
\quad =&& a_0t^{-1/2}+a_1+a_2 t^{1/2}+O(t). \label{hkexp}
\end{eqnarray}
As discussed in \cite{FV}, the heat kernel coefficients $a_0$, $a_1$ and $a_2$ describe one-loop
divergences of the vacuum energy in two dimensions in various regularization schemes. $a_0$ corresponds
to quadratic divergences, $a_1$ -- to linear divergences, and $a_2$ -- to logarithmic ones. 
In a sense, the heat kernel coefficients tell which ``effective geometry" is seen by the quantum fluctuations. For a non-singular
potential, there are analytic expressions for these coefficients, see e.g. \cite{HKrev}. In particular,
the $a_0$ coefficient has to be $(4\pi)^{-1/2}$ times the length of the interval, $\pi/\sqrt{3}$. This
value is consistent with (\ref{hkexp}). The $a_1$ coefficient appears if there are boundaries. For a smooth potential
on an interval with Dirichlet boundary conditions at both endpoints it has to be $-1/2$, which differs from
the value in (\ref{hkexp}). In the nonsingular case $a_2$ is proportional to the integral of the potential.
In our case this integral is divergent, so that no comparison is possible. Generally speaking, there is
no theory of the heat kernel expansion for singular potentials like \eqref{U1}. As well, there are no general
methods of renormalization. Developing such methods is an interesting problem, which we shall not
address here. We shall use the minimal subtraction scheme, that does not require precise understanding
of the counterterms.

In the zeta-function regularization only the logarithmic divergence shows up, that becomes a pole $1/s$.
By using (\ref{Ehk}) one can easily show that the pole contribution to the vacuum energy reads \cite{FV}
\begin{equation}
\mathcal{E}_{\rm pole}=-\frac{a_2}{4\sqrt{\pi}} \, \frac 1s .\label{Epole}
\end{equation}

Let us now represent the sum \eqref{Ezeta} as an integral in the complex plane
\begin{equation}
\mathcal{E}_s=\oint \frac {dy}{4i} (3(y^2+4y))^{\frac 12 -s}{\rm ctg}\, (\pi y)\,,
\label{oint}
\end{equation}
where the contour goes anti-clockwise around the poles of ${\rm ctg}\, (\pi y)$ at the points
$y=1,2,3,\dots$. Next, we rotate the upper part of the contour to the vertical line
$y=\tfrac 12 +iw$, and the lower part -- to $y=\tfrac 12 -iw$, $w\ge 0$; we get 
\begin{eqnarray}
&&\mathcal{E}_s=-\frac 14\! \int_0^\infty \!\!dw\, \tanh(\pi w) \Bigl( e^{-i\pi s} [ 3(w^2\!-\tfrac94\! -5iw)]^{\frac12-s}
\nonumber\\
&&\qquad\qquad +\, e^{i\pi s} [ 3(w^2-\tfrac 94 +5iw)]^{\frac 12 -s} \Bigr). \label{rot}
\end{eqnarray}
To evaluate this integral, we represent $\tanh(\pi w)=1+( \tanh(\pi w)-1)$. The integral of the first (constant)
term is divergent at $s=0$. It has to be evaluated analytically, yielding a pole $1/s$ and a finite contribution.
The integral of $( \tanh(\pi w)-1)$ is finite at $s=0$, and we can put $s=0$ there immediately. After the
calculations, we have for $s\sim 0$:
\begin{equation}
\mathcal{E}_s\simeq -\frac {\sqrt{3}}{2s} + 2.29822\dots \label{num}
\end{equation}
The pole part of this expression is consistent with (\ref{Epole}). As announced above, we use the minimal
subtraction scheme. Namely, we subtract from $\mathcal{E}_s$ the pole (\ref{Epole}) and nothing more.
The renormalized value for the vacuum energy thus becomes
\begin{equation}
\mathcal{E}_{\rm ren}=(\mathcal{E}_s-\mathcal{E}_{\rm pole})_{s=0}=2.29822\dots\label{Efin}
\end{equation}
Since the spectrum of quantum fluctuations is known explicitly, see \eqref{uu} and \eqref{omegan}, one can immediately
construct the propagators and calculate quantum corrections to other quantities of interest.
\section{Comments and conclusions}\label{sec-com}

In this work we have investigated quantum corrections to the mass of a compacton in one of the 
k-field models. This quantum correction is well defined. In general, quantum theory on the background of a compacton does not look more complicated than the one above regular solitons. A complete renormalization theory is still missing, but this is rather a feature of all k-field models than a specific drawback of compactons. 

In conclusion, compactons provide a natural mechanism of confinement and can be successfully quantized. They definitely deserve more attention. 

As a particularly interesting route, we recall that in the recent work \cite{b} one shows how to get from kinks to compactons in models with standard kinematics. This route seems to provide an alternative way to investigate how the quantum corrections behave in the limit where the kinklike solution tends to become a compacton.

\section*{Acknowledgments}

The authors would like to thank CNPq and FAPESP for partial financial support.


\end{document}